\documentclass[12pt]{article}

\usepackage{algorithm}
\usepackage{algorithmicx}
\usepackage{algpseudocode}
\usepackage{amsmath}
\usepackage{amsfonts}
\usepackage{amsthm}
\usepackage{pifont}
\usepackage{threeparttable}
\usepackage{graphicx} 
\usepackage{wrapfig} 
\usepackage{natbib}
\usepackage{setspace}
\usepackage{amssymb}
\usepackage{booktabs}

\newcommand{\argmax}{\operatornamewithlimits{argmax}}

\renewcommand{\vector}[1]{\mathbf{#1}}

\DeclareMathOperator{\logit}{logit}

\graphicspath{{figures/}}

\title{Time-Varying Gaussian-Cauchy Mixture Models for Financial Risk Management}
\author{Shuguang Zhang, Minjing Tao\footnote{To whom correspondence should be addressed: Minjing Tao (tao@stat.fsu.edu).}, Xu-Feng Niu, and Fred Huffer
\\ Department of Statistics, Florida State University}

\begin{document}
\maketitle

\begin{abstract}
There are various metrics for financial risk, such as value at risk (VaR), expected shortfall, expected/unexpected loss, etc. When estimating these metrics, it was very common to assume Gaussian distribution for the asset returns, which may underestimate the real risk of the market, especially during the financial crisis. 
In this paper, we propose a series of time-varying mixture models for risk analysis and management. These mixture models contain two components: one component with Gaussian distribution, and the other one with a fat-tailed  Cauchy distribution.
We allow the distribution parameters and component weights to change over time to increase the flexibility of the models. 
Monte Carlo Expectation-Maximization algorithm is utilized to estimate the parameters. To verify the good performance of our models, we conduct some simulation studies, and implement our models to the real stock market. 
Based on these studies, our models are appropriate under different economic conditions, and the component weights can capture the correct pattern of the market volatility.

\vspace{0.2in}
\end{abstract}

\newpage

\section{Introduction}

Financial market is a place where excessive savings from various resources are mobilized towards those who need them, and also a place where people trade financial resources. Key functions of financial market include borrowing and lending, determination of prices, liquidity, risk sharing, etc. In financial markets, different investments involve different levels of risks, and financial risk management plays an important role in the operation of financial markets. Statistical and mathematical models have been widely used to quantify risks in financial markets and help people avoid large losses.

Year 2008 witnessed a devastating financial crisis which is considered by many economists as the worst financial crisis after the great recession of the 1930s. 
With the burst of house price bubble, the whole financial industry was severely affected, and the crisis in financial sectors then subsequently caused the crash of stock market and started over four-year recession of the whole US economy.
During and after the financial crisis, some statistical models and tools were criticized since they failed to predict the upcoming crisis and financial institutes did not have enough funding to deal with massive amount of delinquencies. 

Among the various financial risk metrics such as value at risk (VaR), expected shortfall (ES, also known as conditional VaR), economic capital (EC), expected and unexpected loss (EL/UL),  VaR is a widely used one, and also faces many criticisms. People even proposed to ban VaR after the 2008 financial crisis.
One of the reasons that VaR failed during the crisis is that asset returns are assumed to have Gaussian distribution when estimating VaR. Since Gaussian distribution has a relatively thin tail, this method can substantially underestimate the market risks, especially when a crisis occurs. 

To overcome this issue, an alternative distribution for the asset returns should be considered. Distributions with larger kurtosis (or say fat-tailed or heavy-tailed distributions) are one of the options, such as student's $t$, Gumbel, Levy, and Cauchy distributions \citep{ferguson1978maximum}. Pure fat-tailed distributions can avoid the problem of risk underestimation, but are sometimes conservative. A class of mixture models can be a better choice.

One of the solutions is to use Gaussian mixture models. \cite{kon1984models} proposed a mixture models of Gaussian distributions to the asset returns and applied the model to 30 stocks in the Dow-Jones Industrial Average. \cite{venkataraman1997value} utilized a mixture model with two mixture components to construct the VaR measures. See also \cite{wirjanto2009applications, rezek2011constrained}. These models can provide a reasonable fit for the data. However, the parameters of the distributions, as well as the weights of individual distributions, were assumed to be constants. 

In reality, financial markets can be very dynamic, and time-varying mixture models are needed to capture this dynamics. 
\cite{wong2000mixture} proposed a mixture autoregressive (MAR) model with $K$ components, where each component followed a Gaussian distribution with its mean having an autoregressive structure. Motivated by the autoregressive conditional heteroscedasticity (ARCH) type models \citep{engle1982autoregressive}, \cite{wong2001mixture} extended their model to the mixture autoregressive conditional heteroscedastic (MAR-ARCH) model. This model consisted of $K$ autoregressive components with conditional heteroscedasticity, creating both time-varying means and standard deviations. \cite{zhang2006mixture} further extended the model to a mixture generalized autoregressive conditional heteroscedastic (M-GARCH) model. 
Besides the time-varying distribution parameters, the weights of each distributions may depend on the time. \cite{wong2001logistic} proposed a logistic mixture autoregressive with exogenous variables (LMARX) model, where the weights of the component follow a logistic regression model with exogenous variables as predictors.

On the other hand, Gaussian distribution is not the only choice for the mixture models.
\cite{li2012cauchy} proposed a Cauchy-Gaussian mixture model for financial risk management. Compared to the Gaussian mixture models, this model captured the distribution of asset returns more accurately, but the parameters in the model were all assumed to be constant over time. Both Gaussian and Cauchy distributions are examples of stable distributions \citep{fama1971parameter}, while Cauchy distribution can further result in a fat-tail \citep{fama1968some}. In terms of mixture, one may also take generalized extreme value (GEV) distributions into consideration. GEV is a family of distributions used to model the maxima and minima of observations, including Fr\'{e}chet, Weibull and Gumbel distributions.

In this paper, we develop a series of mixture models with time-varying parameters. Starting with a model which is assumed to be independent and has constant parameters within each time interval, we then relax the assumption of independency, and reduce the number of parameters in our second model. In this model 2, we consider the location and scale parameters of distributions keep as constants through all the time intervals, and only allow the weighting coefficient to be time-varying across the time intervals. 
In this case, the time-varying weights will become a better indicator to the market risk. We then utilize exogenous variables to model and thus possibly predict the time-varying weights in model 3, and further consider the temporal correlation among weights in model 4. 

The rest of paper proceeds as follows. Section 2 setups the model, with detailed estimation procedures and the corresponding algorithms stated in section 3. We conduct the simulation studies in section 4, and implement our models to the real data in section 5. Section 6 concludes the whole paper.  

\section{Model Set-up}
In this section, we first propose a general class of two-component mixture models with time-varying parameters. We then develop a series of specific models by imposing different assumptions and constraints on the parameters. 
A class of two-component time-varying mixture (TVM) models can be formulated as follows:
\begin{equation}\label{TVM}
f(y_t, \Theta_t\vert \mathcal{F}_{t-1})=\alpha_t f_1(y_t, \Theta_{1, t}) + (1-\alpha_t)f_2(y_t, \Theta_{2, t}),\quad t = 1, 2, \dots, n,
\end{equation}
where $y_t$ is the observed data, $f_1$ and $f_2$ are probability density functions of the two components with parameters $\Theta_{1, t}$ and $\Theta_{2, t}$, respectively. The weight of the first component, $\alpha_t$, is allowed to change over time. 

This TVM model can be considered as a extension of many earlier mentioned models. For example, if $\alpha_t$, $\Theta_{1, t}$ and $\Theta_{2, t}$ are set to be constants, Eq. (\ref{TVM}) is a regular two-component mixture model. Also, all of the MAR, MAR-ARCH, M-GARCH models can be considered as special cases of two-component TVM model with constant $\alpha_t$ and parameters in the Gaussian distributions having time series structure. In addition, LMARX model also has the TVM structure with two components $f_1$, $f_2$ are Gaussian distributions having autoregressive means, and the weight parameter $a_t$ following a logistic model with exogenous variables as predictors. 

The TVM model we will further discuss below consists of two types of distributions: one distribution has thin tails, such as Gaussian distribution, which represents the behavior under the good economic conditions; the other one has a fat tail, such as Cauchy distribution. During the stressed periods, a fat-tailed distribution can help to make conservative decision, and thus prevent from big loss. The time-varing weight $\alpha_t$ helps our model to be suitable for all economic situations, and is an effective indicator of the financial market volatility.

Now we will propose a series of TVM models under different settings and assumptions. 

\paragraph{Model 1: Independent Mixture Models within Time Intervals.}
Rewrite the observed data $\vector{y} = (y_1, \cdots,y_n)$ as $\vector{y} = (y_{11}, \cdots,y_{1n_1},y_{21}, \cdots,y_{2n_2}, y_{k1}, \cdots,y_{kn_k})$, where sample size $n = n_1 + \cdots + n_k$. This idea is to regroup the data according to $k$ time intervals, for example $k$ years, and the number of observations within each time interval is $n_k$. In this sense, $y_{ij}$ represents the $j^{th}$ observation in the $i^{th}$ time interval. In each time interval, we assume that the distribution parameters $\Theta$'s and weight $\alpha$'s are constant, and they are independent with each other. More specifically, our first model is that for $i = 1, 2, \dots, k,\ j = 1, \cdots, n_i$,
\begin{eqnarray}
f(y_{ij}\vert \mathcal{F}_{i,j-1}) &=& \alpha_i f_g(y_{ij}; \mu_i, \sigma_i) + (1-\alpha_i)f_c(y_{ij}; \theta_i, \delta_i),  \label{eq:gcmixmodel}\\
f_g(y_{ij}; \mu_i, \sigma_i) &=&\frac{1}{\sqrt{2\pi}\sigma_t}\exp\left(-\frac{(y_{ij}-\mu_i)^2}{2\sigma_i^2}\right), \nonumber\\
f_c(y_{ij}; \theta_i, \delta_i)&=&\frac{1}{\pi\delta_i}\left( 1+\left(\frac{y_{ij}-\theta_i}{\delta_i}\right)^2 \right)^{-1}, \nonumber
\end{eqnarray}
where $f_g$ and $f_c$ are the densities of Gaussian and Cauchy distributions, respectively, and distributions $f_g$'s and $f_c$'s are all independent with each other. Please notice that although we only mention Cauchy distribution in Eq. (\ref{eq:gcmixmodel}), other fat-tailed distributions maybe used as well, such as student's $t$, Gumbel, and Levy distributions. Similar for the rest three models, we will only use Cauchy distribution to demonstrate the idea.

The idea of Eq. (\ref{eq:gcmixmodel}) is simple: we may separate the financial data according to years, and within each year we estimate the corresponding parameters and the weight. The estimation process of this model is relative easy, but one should understand that this model is actually to treat the data in each time interval separately, and ignore the potential relationships among time intervals.  

\paragraph{Model 2: Mixture Model with Constant Distribution Parameters.}
Model 1 involves a large number of parameters to be estimated, which may not be necessary. Hence in the second model, we assume the distribution parameters of the two density functions are constants, and only let the weighting parameter $a_i$ to be time-varying.  With this assumption, we regard all the data are actually from the same two distributions, but the weight of the Gaussian distribution in each time interval will change through the time. Specifically, the model is stated as follows: for $i = 1, 2, \cdots, k$, $j = 1,2,\cdots,n_i$,
\begin{equation}
	\label{eq:gcmixture2}
f(y_{ij}\vert \mathcal{F}_{i,j-1}) = \alpha_i f_g(y_{ij}; \mu, \sigma) + (1-\alpha_i)f_c(y_{ij}; \theta, \delta).
\end{equation}

In this model, we do not let all the parameters $(\mu, \sigma, \theta, \delta, \alpha)$ change through the time to represent the change of the market. Instead, the fixed two distributions $f_g(\mu,\sigma)$ and $f_c(\theta,\delta)$ are considered as two market situations: low risk and high risk markets, and the only time-varying parameter $\alpha_i$ becomes an indicator of the market risk. By using the model in Eq. (\ref{eq:gcmixture2}), the number of parameters drops from $5k$ to $(k+4)$ for $k$ time intervals. This change will make the model to be more interpretable.

\paragraph{Model 3: Mixture Model with Exogenous Variables.}
To further relaxes the restriction that the weight $\alpha$ is a constant in each time interval, we use exogenous variables as predictors to estimate the time-varying weights in our model 3. The location and scale parameters of Gaussian and Cauchy distributions are still constants. The model is expressed as follows: for $t = 1, 2, \cdots, n$,
\begin{eqnarray}
	f(y_t\vert \mathcal{F}_{t-1}) &=& \alpha_t f_g(y_t; \mu, \sigma) + (1-\alpha_t)f_c(y_t; \theta, \delta), \label{eq:gcmixture3}\\
	\alpha_t &=& \frac{\exp(\vector{x_t}^T\pmb{\beta})}{1+\exp(\vector{x_t}^T\pmb{\beta})}\label{eq:alpha},
\end{eqnarray}
where $\vector{x_t}$ is an exogenous predictor vector with macroeconomic indexes, and $\pmb{\beta}$ is their coefficients. With Eq. (\ref{eq:alpha}), model 3 can also be used to predict the value of $\alpha_t$.

Notice that model 3 can be also applied to the situation that the response $y_t$ and predictors $\vector{x_t}$ are sampled at different frequencies. For example, if $y_t$ is the weekly data, and $\vector{x_t}$ is the monthly data, then model 3 can be written as for $i = 1, 2, \cdots, k$, $j = 1,2,\cdots,n_i$,
\begin{equation}
\label{eq:gcmixture3extra}
\begin{aligned}
	f(y_{ij}\vert \mathcal{F}_{i,j-1}) &=  \alpha_i f_g(y_{ij}; \mu, \sigma) + (1-\alpha_i)f_c(y_{ij}; \theta, \delta), \\
	\alpha_i &= \frac{\exp(\vector{x_i}^T\pmb{\beta})}{1+\exp(\vector{x_i}^T\pmb{\beta})},
\end{aligned}
\end{equation}
where $n_i = 4$ or 5, depending on the number of weeks in a month.

\paragraph{Model 4: Mixture Model with Temporal Correlation.}
Our last model utilizes a logistic dynamic regression model for the weights $\alpha_t$ to incorporate the temporal correlation structure. The model is constructed as follows: for $t = 1, 2, \dots, n$,
\begin{eqnarray}
	f(y_t \vert \mathcal{F}_{t-1}) &=&\alpha_t f_g(y_t; \mu, \sigma) + (1-\alpha_t)f_c(y_t; \theta, \delta), \label{eq:gcmixture4} \\
	\alpha_t &=& \frac{\exp(\vector{x_t}^T\pmb{\beta}+e_t)}{1+\exp(\vector{x_t}^T\pmb{\beta}+e_t)}, \nonumber\\
	e_t & =& \phi_1 e_{t-1} + a_t, \quad a_t \overset{i.i.d.}{\sim}N(0, \sigma_a^2), \label{eq:ar1error} \nonumber
\end{eqnarray}
where $e_t$ is an autoregressive time series AR(1). If we set the coefficients $\pmb\beta$ to be zeros, the model for weight $\alpha_t$ is simply a logistic function of an AR time series which can model the correlation structure and have an appropriate range (0, 1). In addition, the model remains valid under the case with different sampling frequencies of $y_t$ and $\vector{x_t}$. One just need to re-write the model to the similar format in Eq. (\ref{eq:gcmixture3extra}).

\section{Estimation Procedure and Algorithms}
The main technique we will use for estimation is the EM algorithm, which has been used as a classical method to estimate parameters in mixture models. This algorithm goes through expectation (E-step) and maximization (M-step), and iteratively update parameters until convergence. Theoretical proof of guaranteed global optimal can be found in \cite{dempster1977maximum}. 

In our models, the observed data are $\vector{y}=(y_{11}, \dots, y_{1n_1}, y_{21}, \dots, y_{2n_2}, y_{k1},\dots, y_{kn_k})^T$, and $\vector{X}=(\vector{x_1}, \cdots,\vector{x_1}, \vector{x_2}, \dots, \vector{x_2}, \vector{x_k},\cdots,\vector{x_k})^T$\footnote{Here we directly write down the format for different sampling frequencies of $\vector{y}$ and $\vector{X}$. If they have the same frequency, then we have $k = n$ and $n_i = 1$ for all $i = 1, \cdots,n$. In this case, the observed data are $\vector{y} = (y_1, \cdots, y_n)^T$ and $\vector{X} = (\vector{x_1},\cdots, \vector{x_n})$.}, where $y_{ij}$ is the value of $j^{th}$ observed response variable at time $i$ for $i = 1, 2, \dots, k$ and $\vector{x_i}=(x_{i1}, x_{i2}, \dots, x_{ip})^T$ is a $p$-dimensional predictor vector at time $i$ (we only have $\vector{X}$ in model 3 and 4). The unobserved data include $\pmb\alpha = (\alpha_1, \alpha_2, \dots \alpha_k)^T$ and $\vector{z}=(z_{11}, \dots, z_{1n_1}, z_{21}, \dots, z_{2n_2}, z_{k1}, \dots, z_{kn_k})^T$, where $\alpha_i$ is the weight of Gaussian component at time $i$; and $\vector{z}$ is the indicator of the type of the distribution. Specifically, $z_{ij}=1$ if the $j^{th}$ observation at time $i$ is from Gaussian distribution, $z_{ij}=0$ if it is from Cauchy distribution, and $P(z_{ij} = 1) = \alpha_i$.

\subsection{EM Algorithm for Models 1--3}

We will first discuss the parameter estimations for the first three models, whose differences are very small. For the illustration purpose, we only mention the details for the Gaussian-Cauchy mixture model with constant weights and parameters in one time interval as a demonstration.
 
The first step of EM algorithm is to compute expectation of log-likelihood over the missing data based on the current iterates, which can be derived as
\begin{equation}
\setlength\abovedisplayskip{12pt}
\label{eq:expectation}
	\begin{aligned}
	E\left(\log f(\mathbf{y}\rvert\Theta)\rvert \mathbf{y}, \Theta_k\right)
	 & = \sum\limits_{i=1}^{n}\left(\left(\log\alpha - \log\sigma - \frac{(y_i-\mu)^2}{2\sigma^2}\right)p_i \right.\\
	 + & \left. \left(\log(1-\alpha)-\log\delta-\log\left(1+\left(\frac{y_i-\theta}{\delta}\right)^2\right)\right)(1-p_i)\right),
	\end{aligned}
\end{equation}
where $\mathbf{y}=(y_1, y_2, \dots, y_{n_1})^T$ denotes the observed data in one time interval, $\Theta=(\alpha, \mu, \sigma, \theta, \delta)$ is the parameter space, $\Theta_k=(\alpha_k, \mu_k, \sigma_k, \theta_k, \delta_k)$ is its $k$-th iterates, and $p_i$ is the probability that the $i$-th observation is from the Gaussian population, which is calculated by
\begin{equation*}
p_i = \frac{\alpha_k f_g(y_i\rvert \mu_k, \sigma_k)}{\alpha_k f_g(y_i\rvert \mu_k, \sigma_k)+(1-\alpha_k) f_c(y_i\rvert \theta_k, \delta_k)}.
\end{equation*}

To maximize the expectation in Eq. (\ref{eq:expectation}), it is not difficult to see the solutions for $\alpha$, $\mu$, and $\sigma^2$ are
 $$\hat{\alpha}=\frac{1}{n}\sum\limits_{i=1}^{n}p_i,
\quad \hat{\mu}=\frac{\sum\limits_{i=1}^{n}p_i y_i}{\sum\limits_{i=1}^{n}p_i}, 
\quad \hat{\sigma}^2 = \frac{\sum\limits_{i=1}^{n}(y_i - \hat{\mu})^2 p_i}{\sum\limits_{i=1}^{n}p_i}.$$ 
However, there is no closed-form solutions for $\theta$ and $\delta$ in Eq. (\ref{eq:expectation}). Hence, we utilize the Newton-Raphson method to  obtain their optimal estimates. 

Since the main idea to estimate the parameters in the first three models are very similar, we only provide the Algorithm \ref{alg:gcmixture3} for model 3 defined in Eq. (\ref{eq:gcmixture3}) here. The algorithms for the other two models are provided in Appendix.

\begin{algorithm}[t!]
	\caption{Parameter Estimation for Gaussian-Cauchy Mixture Model with Constant Location and Scale Parameters and Exogenous Variables as Predictors}
	\label{alg:gcmixture3}
	\begin{algorithmic}[1]
		\Require{data vector $\mathbf{y}=(y_{11}, \dots, y_{1n_1}, y_{21}, \dots, y_{2n_2}, y_{k1}, \dots, y_{kn_k})^T$, predictor matrix $\mathbf{X}=(\vector{x_1}, \dots, \vector{x_1}, \vector{x_2}, \dots, \vector{x_2}, \vector{x_k},\dots, \vector{x_k})^T$, where $\vector{x_i}=(x_{i1}, x_{i2}, \dots, x_{ip})^T$ for $i = 1, 2, \dots, k$}.
		\State Set the initial estimates
		\State $\mu_0 \gets median(\mathbf{y})$
		\State $\sigma_0 \gets IQR(\mathbf{y})$
		\State $\theta_0 \gets median(\mathbf{y})$
		\State $\delta_0 \gets IQR(\mathbf{y})$
		\State $\pmb{\beta_0}=(0, 0, \dots, 0)^T$
		\While {not converge}
		\State $\alpha_{i0}=\frac{\exp(\vector{x_i}^T\pmb{\beta_0})}{1+\exp(\vector{x_i}^T\pmb{\beta_0})}$ for $i = 1, 2, \dots, k$
		\State  $p_{ij} \gets \frac{\alpha_{i0} f_g(y_{ij}, \mu_0, \sigma_0)}{\alpha_{i0} f_g(y_{ij}, \mu_0, \sigma_0)+(1-\alpha_{i0}) f_c(y_{ij}, \theta_0, \delta_0)}$ for $i = 1, 2, \dots, k$ and $j = 1, 2, \dots, n_k$
		\State 	$\pmb{\beta_{new}} \gets \argmax\limits_{\pmb{\beta}} \sum\limits_{i=1}^{k}\sum\limits_{j=1}^{n_i}\left(\vector{x_i}^T\pmb{\beta}p_{ij}-\log(\exp(\vector{x_i}^T\pmb{\beta})+1)\right)$ 
		\State	$\mu_{new} \gets \frac{\sum\limits_{i=1}^{k}\sum\limits_{j=1}^{n_i}p_{ij} y_{ij}}{\sum\limits_{i=1}^{k}\sum\limits_{j=1}^{n_i}p_{ij}} $,\\
		\State	$\sigma_{new} \gets \sqrt{\frac{\sum\limits_{i=1}^{k}\sum\limits_{j=1}^{n_i}(y_{ij} - \mu_{new})^2 p_{ij}}{\sum\limits_{i=1}^{k}\sum\limits_{j=1}^{n_i}p_{ij}}}$
		\State	$(\theta_{new}, \delta_{new}) \gets \argmax\limits_{\theta, \delta} \sum\limits_{i=1}^{k}\sum\limits_{j=1}^{n_i}\left(\left(-\log\delta-\log(1+(\frac{x_{ij}-\theta}{\delta})^2)\right)(1-p_{ij})\right)$ 
		\State $\pmb{\beta_0} \gets \pmb{\beta_{new}}, \mu_0 \gets \mu_{new}, \sigma_0 \gets \sigma_{new}, \theta_0 \gets \theta_{new}, \delta_0 \gets \delta_{new}$
		\EndWhile \\
		\Return par = $(\mu_0, \sigma_0, \theta_0, \pmb{\beta_0})$, 
	\end{algorithmic}
\end{algorithm}

\subsection{Monte Carlo EM Algorithm for Model 4}

In model 4, the parameters of interest are $\Theta=(\mu, \sigma, \theta, \delta, \pmb\beta, \phi_1, \sigma_a)$, where $\mu, \theta \in \mathbb{R}$, $\sigma, \delta, \sigma_a \in \mathbb{R}^+$, $\phi_1 \in (-1, 1)$ and $\pmb\beta \in \mathbb{R}^{p}$. In this setup, the conditional density function of observed data $\vector{y}$ and $\vector{z}$ are
\begin{eqnarray}
f(\vector{y}\vert \vector{z}, \Theta) &=& \prod\limits_{i=1}^{k}\prod\limits_{j=1}^{n_i} f_g(y_{ij}\vert\Theta)^{z_{ij}}f_c(y_{ij}\vert\Theta)^{1-z_{ij}},  \label{eq:y_z}\\
f(\vector{z} \vert \pmb\alpha, \Theta) &=& \prod\limits_{i=1}^{k}\prod\limits_{j=1}^{n_i} \alpha_i^{z_{ij}}(1-\alpha_i)^{1-z_{ij}}.
\label{eq:z_alpha}
\end{eqnarray}

For the AR(1) model given in Eq. (\ref{eq:ar1error}), we can find that $\mathbf{e}=(e_1, e_2, \dots, e_k)^T\sim N(\textbf{0}, \Sigma)$ with
$\Sigma = (\Sigma_{ij})$, and
\begin{equation*}
\Sigma_{ij} = \frac{\phi_1^{|i-j|}}{1-\phi_1^2} \sigma^2_a.
\end{equation*}
Define $\vector{u}=\vector{X}\pmb{\beta}+\vector{e}$, then we can find that $\vector{u} \sim N(\vector{X}\pmb{\beta}, \Sigma)$ and 
\begin{equation}
\label{eq:weight_dist}
f(\pmb\alpha\vert \Theta)
	=\frac{(2\pi)^{-\frac{k}{2}}\lvert\Sigma\rvert^{-\frac{1}{2}}
	        \exp\left(-\frac{1}{2}(\vector{u} -\vector{X}\pmb\beta)^T\Sigma^{-1}(\vector{u} -\vector{X}\pmb\beta)\right)}
		  {\prod\limits_{i=1}^{k}\alpha_i(1-\alpha_i)},
\end{equation}
where $\vector{u} = (u_1, u_2,\cdots,u_k)^T$, and $u_i = \log \frac{\alpha_i}{1-\alpha_i}$. Thus, the log-likelihood of joint density function of observed and unobserved data is
\begin{equation*}
L = \log f(\vector{y}, \vector{z}, \pmb\alpha \vert \Theta) 
= \log\left(f(\vector{y}\vert \vector{z}, \Theta)f(\vector{z} \vert \pmb\alpha, \Theta)f(\pmb\alpha\vert \Theta)\right), 
\end{equation*}
where the three conditional densities are given in Eq. (\ref{eq:y_z}), (\ref{eq:z_alpha}), and (\ref{eq:weight_dist}). Then in the E-step, the expectation of complete log-likelihood with respect to the conditional density of unobserved data is 
\begin{equation} \label{E-step}
		E(L\vert \vector{y}, \Theta_k)
		= E_{\alpha}\left( E(L\vert \pmb\alpha, \vector{y}, \Theta_k) \vert\vector{y}, \Theta_k\right). \\
\end{equation}
To calculate Eq. (\ref{E-step}), we need to (1) use the fact that
$$E(z_{ij}|\alpha, \vector{y}, \Theta_k) = p_{ij} \triangleq
\frac{\alpha_i f_g(y_{ij}\vert \Theta_k)}{\alpha_i f_g(y_{ij}\vert \Theta_k)+(1-\alpha_i)f_c(y_{ij}\vert \Theta_k)},$$
and (2) evaluate  $f(\pmb\alpha \vert \vector{y}, \Theta_k)$ which is obtained by
\begin{equation*}
 f(\pmb\alpha \vert \vector{y}, \Theta_k) = \frac{f(\pmb\alpha, \vector{y} \vert \Theta_k)}{f(\vector{y}\vert \Theta_k)} 
= \frac{f(\vector{y}\vert \pmb\alpha) f(\pmb\alpha \vert \Theta_k)}{f(\vector{y}\vert \Theta_k)} 
= \frac{f(\vector{y}\vert \pmb\alpha) f(\pmb\alpha \vert \Theta_k)}{\int f(\vector{y}\vert \pmb\alpha) f(\pmb\alpha \vert \Theta_k) d\pmb\alpha}, 
\end{equation*} 
where $f(\pmb\alpha \vert \Theta_k)$ is given in Eq. (\ref{eq:weight_dist}), and
$$f(\vector{y}\vert \pmb\alpha) = \sum_{\vector{z}}f(\vector{y}, \vector{z} \vert \pmb\alpha) = \prod\limits_{i=1}^{k}\prod\limits_{j=1}^{n_i} \left( \alpha_i f_g(y_{ij}\vert \Theta_k) + (1-\alpha_i) f_c(y_{ij}\vert \Theta_k)\right).$$

It is not easy to find the analytic form of $E(L\vert \vector{y}, \Theta_k)$. Thus we use Monte Carlo methods to find an approximation of the expectation. To achieve this, we can generate a series of Markov chain Monte Carlo (MCMC) samples of $\pmb{\alpha_1}, \pmb{\alpha_2}, \dots, \pmb{\alpha_n} \sim f(\pmb\alpha \vert \vector{y}, \Theta_k)$ and compute the sample mean $\frac{1}{n}\sum\limits_{i=1}^nE(L\vert \pmb{\alpha_i}, \vector{y}, \Theta_k)$, which will subsequently be maximized in the M-step. The detailed estimation procedure can be found in Algorithm \ref{alg:MCEM}.

\begin{algorithm}
	\caption{Monte Carlo Expectation Maximization}
	\label{alg:MCEM}
	\begin{algorithmic}[1]
		\Require{response vector $\vector{y}$ and predictor matrix $\mathbf{X}$, which are the same as in Algorithm \ref{alg:gcmixture3}}. 
		\State Set the initial estimates $\hat\Theta = (\mu_0, \sigma_0, \theta_0, \delta_0, \phi_0, \sigma_{a0}, \pmb\beta_0)$
		\State $\mu_0 \gets median(\vector{y})$
		\State $\sigma_0 \gets IQR(\vector{y})$
		\State $\theta_0 \gets median(\vector{y})$
		\State $\delta_0 \gets IQR(\vector{y})$
		\State $\phi_0 \gets 0.5$
		\State $\sigma_{a0} \gets 1$
		\State $\pmb\beta_0 \gets \mathbf{0}$
		\Repeat 
		\State $\Theta \gets \hat{\Theta}$
		\State Use algorithm \ref{alg:metro_within_gibbs} to generate $samples = (\pmb{\alpha_1}, \pmb{\alpha_2}, \dots, \pmb{\alpha_n})$
		\State $\hat\Theta =\argmax\limits_{\Theta}\frac{1}{n}\sum\limits_{i=1}^nE(L\vert \pmb{\alpha_i}, \vector{y}, \Theta_k)$  where $\pmb{\alpha_i} = (\alpha_{i1}, \alpha_{i2}, \dots, \alpha_{it})$
		\Until converge\\
		\Return $\Theta$
	\end{algorithmic}
\end{algorithm}

One of the methods to generate the MCMC samples is Metropolis-Hasting algorithm. It can be easily implemented, but the MCMC samples mix slowly and will affect prediction accuracy, because the sample vector will update simultaneously. Therefore, in this study we implement Metropolis within Gibbs as an alternative. Gibbs sampler is widely adopted in high dimensional sampling studies and it updates one variable at a time based on the conditional density given all the other variables. 

Let $\alpha_i$, $i = 1, 2, \dots, k$,  denote the $i^{th}$ variable and $\pmb\alpha_{-i}$ denote all the variables but the $i^{th}$ one, then its conditional density function is
\begin{equation*}
f(\alpha_i \vert \vector{y}, \Theta, \pmb\alpha_{-i})  
		 = \frac{f(\pmb\alpha \vert \vector{y}, \Theta)}{f(\pmb\alpha_{-i} \vert \vector{y}, \Theta)} 
		 = \frac{f(\pmb\alpha \vert \vector{y}, \Theta)}{\int f(\pmb\alpha \vert \vector{y}, \Theta) d\alpha_i}. 
\end{equation*}
Define $\vector{v}=\vector{u} -\mathbf{X}\pmb\beta_{\Theta}= (v_1, v_2, \dots, v_t)^T$, then we can derive that
\begin{equation} 
\label{eq:gibbs}
f(\alpha_i \vert \vector{y}, \Theta, \pmb\alpha_{-i}) \propto \left\{
\begin{array}{ll}
g_1 \exp\left( - \frac{v_1^2 - 2\phi_1 v_1 v_2}{2\sigma^2_a} \right) & \text{for } i=1,\\
g_t \exp\left( - \frac{v_t^2 - 2\phi_1 v_{t-1}v_t}{2\sigma^2_a} \right) & \text{for } i=t, \\
g_i \exp\left( - \frac{(1+\phi_1^2)v_i^2 - 2\phi_1(v_{i-1}v_i + v_i v_{i+1})}{2\sigma^2_a} \right) & \text{others,} 
\end{array}
\right.
\end{equation}
where $g_i$ is defined as 
$$g_i = \frac{\prod\limits_{j=1}^{n_i}(\alpha_i f_g(y_{ij}\vert \Theta) + (1-\alpha_i)f_c(y_{ik}\vert \Theta))}{\alpha_i(1-\alpha_i)}.$$
With the conditional densities derived in Eq. (\ref{eq:gibbs}), we can generate the Gibbs sampler, and the detailed computational methods can be found in Algorithm \ref{alg:metro_within_gibbs}.

\begin{algorithm}
	\caption{Metropolis within Gibbs Sampling}
	\label{alg:metro_within_gibbs}
	\begin{algorithmic}[1]
		\Require{response vector $\vector{y}$, predictor matrix $\mathbf{X}$, which are the same as in Algorithm \ref{alg:gcmixture3}, and parameter space $\Theta = (\mu, \sigma, \theta, \delta, \phi_1, \sigma_a, \pmb\beta)$}. 
		\State Initialize the Markov chain using uniform random variables $\pmb{\alpha^{(0)}}= (\alpha^{(0)}_1, \alpha^{(0)}_2, \dots, \alpha^{(0)}_t)'$
		\State $\vector{b}=\mathbf{X}\pmb\beta= (b_1, b_2, \dots, b_t)$
		\State $\vector{v}=\log\frac{\pmb\alpha}{\vector{1}-\pmb\alpha}-\vector{b}= (v_1, v_2, \dots, v_t)$
		\For {$i$ in $1:n$}
		\For {$j$ in $1:t$}
		\State $\logit(\alpha_{new}) \sim N(\logit(\alpha^{(i-1)}_j), \sigma_{rw}^2)$
		\State $v_{new}=\alpha_{new}-b_j$
		\If{$j=1$}
			\State $\rho \gets \min\left(1, \frac{g_{new}
			\exp\left( -(v_{new}^2-2\phi_1v_{new}v_2)/2\sigma_a^2\right)}{ g_1^{i-1}\exp\left(-(v_1^2-2\phi_1v_1v_2)/2\sigma_a^2 \right)}\right)$
		\ElsIf{$j = t$}
			\State $\rho \gets \min\left(1, \frac{ g_{new}
			\exp\left(-(v_{new}^2-2\phi_1v_{new}v_{t-1})/2\sigma_a^2\right)}{ g_t^{i-1}\exp\left(-(v_t^2-2\phi_1v_{t-1}v_t)/2\sigma_a^2 \right)}\right)$
		\Else
			\State $\rho \gets \min\left(1, \frac{ g_{new}
			\exp\left(-(1+\phi_1^2)v_{new}^2-2\phi(v_{j-1}v_{new}+v_{new}v_{j+1})/2\sigma_a^2 \right)}
		{ g_j^{i-1}\exp\left(-(1+\phi_1^2)v_j^2-2\phi(v_{j-1}v_j+v_jv_{j+1})/2\sigma_a^2 \right)}\right)$
		\EndIf
		\State $u \sim unif(0, 1)$
		\If{$u\leq\rho$}
		\State $\alpha^{(i)}_j=\alpha_{new}$
		\State $v_j = v_{new}$
		\Else
		\State $\alpha^{(i)}_j=\alpha^{(i-1)}_j$
		\EndIf
		\EndFor
		\State $\pmb{\alpha^{(i)}}=(\alpha^{(i)}_1, \alpha^{(i)}_2, \dots, \alpha^{(i)}_t)'$
		\EndFor\\
		\Return $samples = (\pmb{\alpha^{(1)}}, \pmb{\alpha^{(2)}}, \dots, \pmb{\alpha^{(n)}})$
	\end{algorithmic}
\end{algorithm}

\section{Simulation Studies}
In this section, we conduct some simulation studies on our mixture models. 
There are multiple ways to generate samples that follow a mixture distribution with density having the structure $f(x) = \alpha f_1(x) + (1 - \alpha)f_2(x)$. 
What we use in this study is to generate $\alpha*N$ random variables with density $f_1$ and $(1 - \alpha)*N$ random variables with density $f_2$. 

\paragraph{Model 1: Independent Models within Time Intervals.}
For the first model defined in Eq. (\ref{eq:gcmixmodel}), we have fixed the two distributions as $N(0, 1)$ and Cauchy$(0, 1)$, and compared the results for different $\alpha$ values and sample sizes $n$. 500 replicates are run in each setting. The results are provided in table \ref{sim:gcmixture}.

The upper panel of table \ref{sim:gcmixture} has the same $\alpha$ value with different sample sizes $n$. It can be seen that the MSEs of $\mu$,  $\sigma$, $\theta$, and $\alpha$ are very small. On the other hand, the mean estimate of $\delta$ is lower than its true value, which indicates a systematic underestimation of the scale parameter in Cauchy distribution. This is potentially due to the large value of $\alpha$, that is, 90\% of the data is from the thin-tailed Gaussian distribution. Hence the estimated $\alpha$ is slightly less than its true value, and thus the scale parameter of Cauchy distribution is underestimated. In addition, when the number of samples is reduced from 100 to 50, the results almost remain the same with slightly higher standard errors. Thus in the simulation studies of the other three models, we will generate samples with $n = 50$.

The upper panel of table \ref{sim:gcmixture} represents the good economic situation that the Gaussian component is dominant. When the economic situation gets bad, the weight of Cauchy component will become larger. Hence we decrease the $\alpha$ value in the lower panel of table \ref{sim:gcmixture}. With $\alpha = 0.5$, meaning that the two distributions will occur equally likely, the estimation is quite accurate, and the underestimation of $\delta$ observed in the upper panel is improved. When $\alpha = 0.1$, one can observe the opposite results: the estimated weight $\hat\alpha$ is towards to the Gaussian distribution, and the scale parameter of Cauchy is overestimated. 

\begin{table}[H]
	\caption{Simulation results for Gaussian-Cauchy mixture model defined in Eq. (\ref{eq:gcmixmodel}). }
	\label{sim:gcmixture}
	\vspace{5pt}
	\centering
	\begin{threeparttable}
		\begin{tabular}{ c || c  c  c ||  c  c  c}
			Parameter & True & Estimate Mean (SE) & MSE & True & Estimate Mean (SE) & MSE\\ [2pt]
			\hline 
			& \multicolumn{3}{|c||}{$n = 100$} & \multicolumn{3}{c}{$n = 50$} \\
			$\mu$ 	 & 0 	& 0.00027 		(0.11) 	& 0.012 
			 	 & 0 	& 0.0036 	(0.16)	& 0.025 \\ [2pt]
			$\sigma$ & 1 	& 1.02 		(0.084) 	& 0.0074 
			 & 1 	& 1.00 		(0.12) 	& 0.015\\ [2pt]
			$\theta$ & 0 	& -0.00088 	(0.13) 	& 0.017 
			 & 0 	& -0.0031 	(0.19) 	& 0.035\\ [2pt]
			$\delta$ & 1 	& 0.64 		(0.064) & 0.14 
			& 1 	& 0.63 		(0.089)  & 0.15\\ [2pt]
			$\alpha$ & 0.9 	& 0.87 		(0.10) 	& 0.010 
			 & 0.9 	& 0.86 		(0.14) 	& 0.021 \\[2pt]
			\hline
			 & \multicolumn{3}{|c||}{$n = 100$} & \multicolumn{3}{c}{$n = 100$} \\
			
					$\mu$ 	 & 0 	& 0.0010 	(0.16)	& 0.026 
					& 0 	& 0.0010 	(0.40)	& 0.16\\ [2pt]
		$\sigma$ & 1 	& 1.02 		(0.19)  & 0.036 
		& 1 	& 0.88 		(0.43)  & 0.20\\ [2pt]
		$\theta$ & 0 	& -0.0031 	(0.12) 	& 0.015 
		& 0 	& 0.0096 	(0.16) 	& 0.025\\ [2pt]
		$\delta$ & 1 	& 0.88 		(0.12) & 0.028
		& 1 	& 1.26 		(0.22)  & 0.11 \\ [2pt]
		$\alpha$ & 0.5 	& 0.47 		(0.11) 	& 0.013 
		& 0.1 	& 0.29 		(0.091)	& 0.045\\[2pt] \hline
		\end{tabular}
	\end{threeparttable}
\end{table}

\begin{figure}[H]
	\begin{center}
		\includegraphics[scale = 0.3]{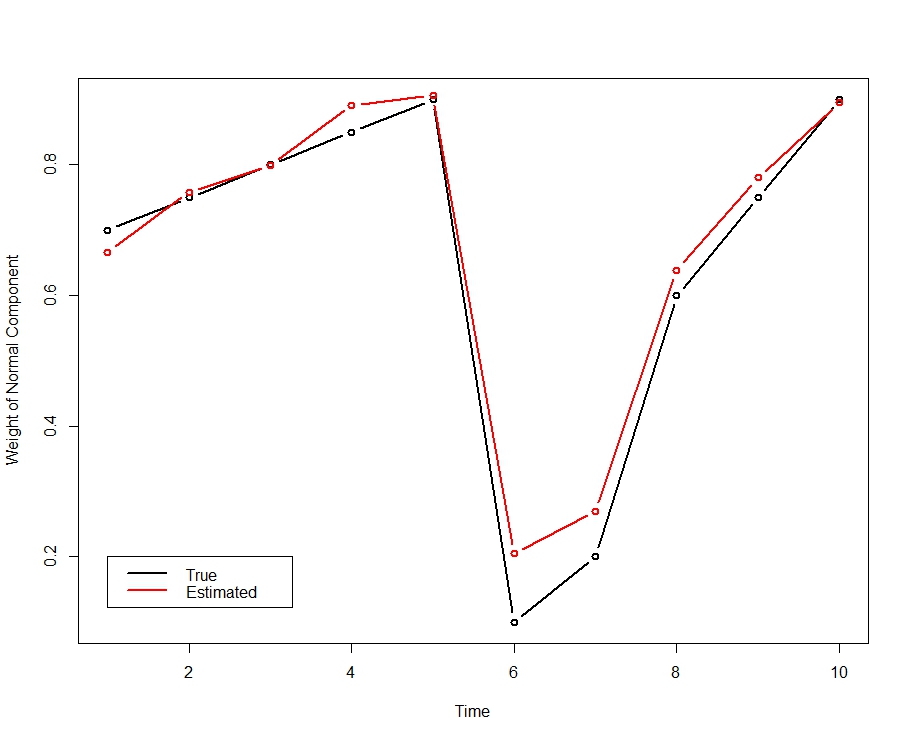}
	\end{center}
	\caption{Estimation results of the weights in different time intervals based on model in Eq. (\ref{eq:gcmixture2}). The detailed information is provided in table \ref{sim:m2gcmixture_extra} in Appendix.}
	\label{fig:model2}
\end{figure}

\paragraph{Model 2: with Constant Distribution Parameters.}
For the second model proposed in Eq. (\ref{eq:gcmixture2}), we still use the fixed two distributions $N(0, 1)$ and Cauchy$(0, 1)$, and also for the rest two models. We apply model 2 to 10 time intervals with 50 observations in each time interval. The weights $\alpha_i$ vary among different time intervals. We repeat this simulation process for 100 times. The estimation results are shown in table \ref{sim:m2gcmixture} and figure \ref{fig:model2}. 

\begin{table}[H]
	\caption{Simulation results for Gaussian-Cauchy mixture model proposed in Eq. (\ref{eq:gcmixture2}).}
	\label{sim:m2gcmixture}
	\vspace{5pt}
	\centering
	\begin{threeparttable}
		\begin{tabular}{ c | c  c  c  c }
			Parameter & True Value & Estimate Average & Estimate SE & MSE \\ [2pt]
			\hline 
			$\mu$ 	 & 0 	& -0.000040 	& 0.063	& 0.0039 \\ [2pt]
			$\sigma$ & 1 	& 0.99 			& 0.050 & 0.0026 \\ [2pt]
			$\theta$ & 0 	& -0.0020		& 0.11 	& 0.0126 \\ [2pt]
			$\delta$ & 1 	& 1.08 			& 0.11  & 0.0197 \\ [2pt]
			\hline
		\end{tabular}
	\end{threeparttable}
\end{table}

From table \ref{sim:m2gcmixture}, the estimation results of distribution parameters are very accurate. In this simulation, we have $50\times10=500$ observations to estimate the distribution parameters in each replication. Thus although the weights $\alpha_i$ may be away from 0.5, there is no underestimation nor overestimation for the scale parameter $\delta$. In addition, based on figure \ref{fig:model2}, the estimated weights $\hat\alpha_i$ can capture the true changing pattern of $\alpha_i$.

We need to point out that there will involve a large number of weights $\alpha_i$ to be estimated in model 2, especially when the number of time intervals is large. In addition, there is not much predictive power to $\alpha_i$ from this model. Thus we will move to the next model in which the weights can be expressed as a logistic function of exogenous variables.

\begin{figure}[H]
	\begin{center}
		\includegraphics[scale = 0.22]{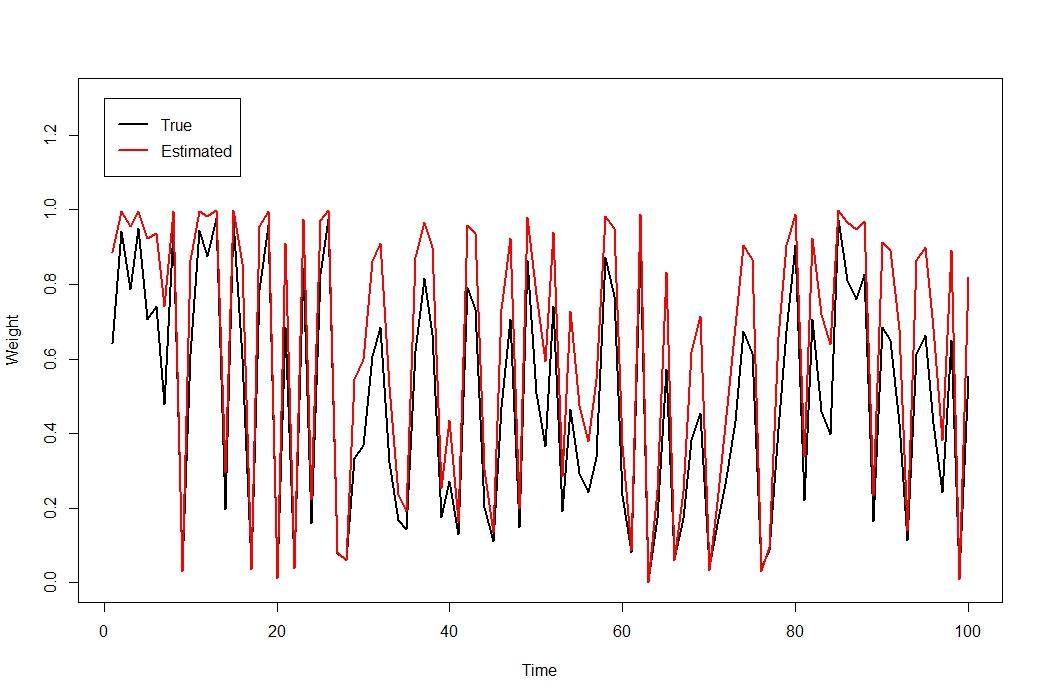}
		\includegraphics[scale = 0.22]{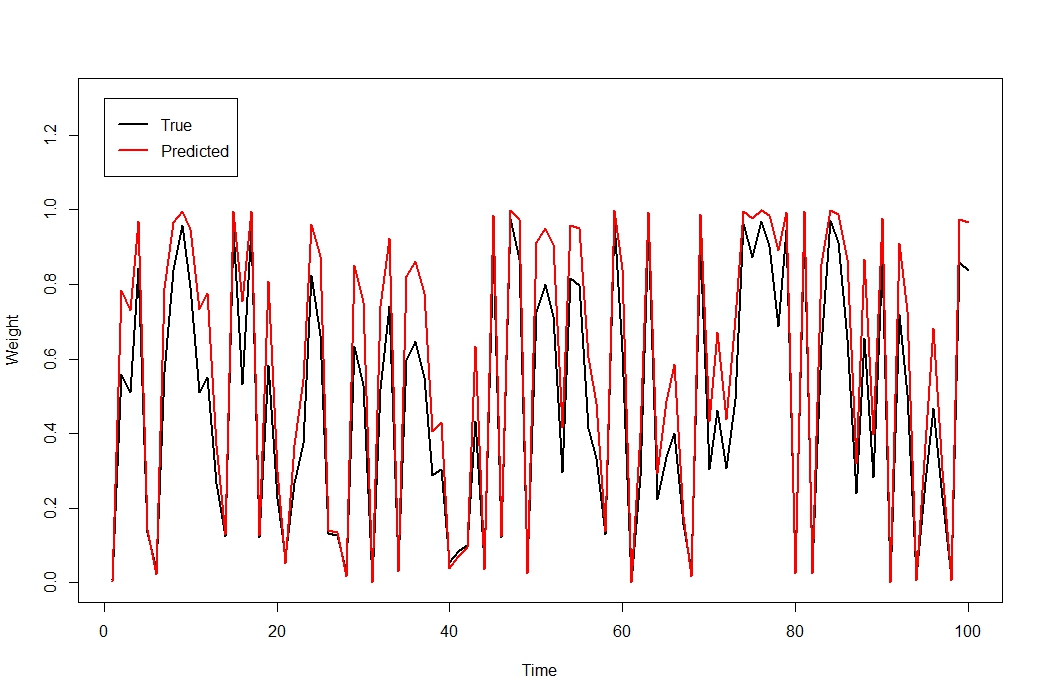}		
	\end{center}
	\caption{Weights estimation/prediction results for the model defined in Eq. (\ref{eq:gcmixture3}) with six exogenous variables. Left plot: estimated weights under training data; right plot: predicted weights under test data.}
	\label{fig:model3_p6}
\end{figure}

\paragraph{Model 3: with Exogenous Variables.}
In this simulation, we increase the number of time intervals to 100, while the sample size in each interval remains 50. The whole process is repeated 100 times. To study the effects of the number of variables, we conduct the simulation for $p = 1, 3, 6$. The results are very similar to each other. Hence we only present the results for $p = 6$ in table \ref{sim:model3_p6} and figure \ref{fig:model3_p6}.

From table \ref{sim:model3_p6} we can find that the estimation results for the two distributions are very accurate, similar to table 2. On the other hand, the estimations for $\pmb\beta$ may be far away from the truth. However, this difference will not affect the model ability to capture the correct changing pattern of $\alpha_i$ (see figure \ref{fig:model3_p6}).

In order to check the predictive power of model 3, we first estimate the coefficients $\pmb\beta$ for exogenous variables, and thus obtain the weight estimates $\hat\alpha_i$ based on the training dataset. Then we apply the fitted model to the test dataset and calculate the predicted weights. From figure \ref{fig:model3_p6}, it can be observed that besides the estimation, our model 3 can also be used to predict the future weights, which may be used as an indicator of the market volatility. The overall MSEs of the weights from the training and test sets are 0.08133 and 0.08679, respectively.  

\begin{table}
	\caption{Simulation results for the model defined in Eq. (\ref{eq:gcmixture3}) with six exogenous variables.} 
	\label{sim:model3_p6}
	\vspace{5pt}
	\centering
	\begin{threeparttable}
		\begin{tabular}{ r | c  c  c  c }
			Parameter & True Value & Estimate Average & Estimate SE & MSE \\ [2pt]
			\hline 
			$\mu$ 	 & 0 	& -0.0017 	& 0.018		& 0.00032 \\ [2pt]
			$\sigma$ & 1 	& 1.05 		& 0.018 	& 0.0028 \\ [2pt]
			$\theta$ & 0 	& -0.0031  	& 0.044 	& 0.0019 \\ [2pt]
			$\delta$ & 1 	& 1.31 		& 0.088  	& 0.010 \\ [2pt]\hline
			$\beta_0$ & 1	& 3.57 		& 0.20 		& 6.67 \\[2pt]
			$\beta_1$ & 2	& 1.15 		& 0.24 		& 0.78 \\[2pt]
			$\beta_2$ & -3	& -2.12 	& 0.18 		& 0.80 \\[2pt]
			$\beta_3$ & 1	& 0.66 		& 0.22 		& 0.16 \\[2pt]
			$\beta_4$ & 4	& 2.78 		& 0.21 		& 1.54 \\[2pt]
			$\beta_5$ & 3	& 2.06 		& 0.19 		& 0.92 \\[2pt]
			$\beta_6$ & 0.5	& 0.72 		& 0.26 		& 0.12 \\
			\hline
		\end{tabular}
	\end{threeparttable}
\end{table}

\paragraph{Model 4: When Temporal Correlation Considered.}
For model 4 given in Eq. (\ref{eq:gcmixture4}), we simulate data from 100 time intervals with 50 observations each, and use two exogenous variables as predictors. MCEM method is used for the estimation. Due to the large sample size and complicated model formulation, it takes long time (hours to days) to obtain the estimation results. In addition, due to the randomness in the MC procedure, the estimation results may be unstable, especially when the weights $\alpha_i$ change drastically through the time. However, the good news is that our model 3 still works when there actually exists temporal correlations in $\alpha_i$.

\begin{figure}
	\begin{center}
		\includegraphics[scale = 0.25]{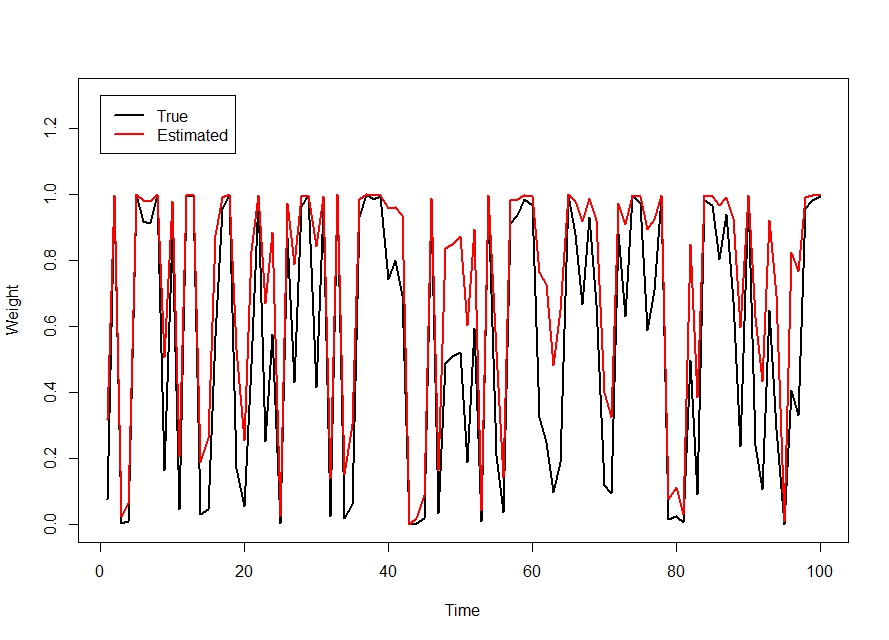}
		\includegraphics[scale = 0.25]{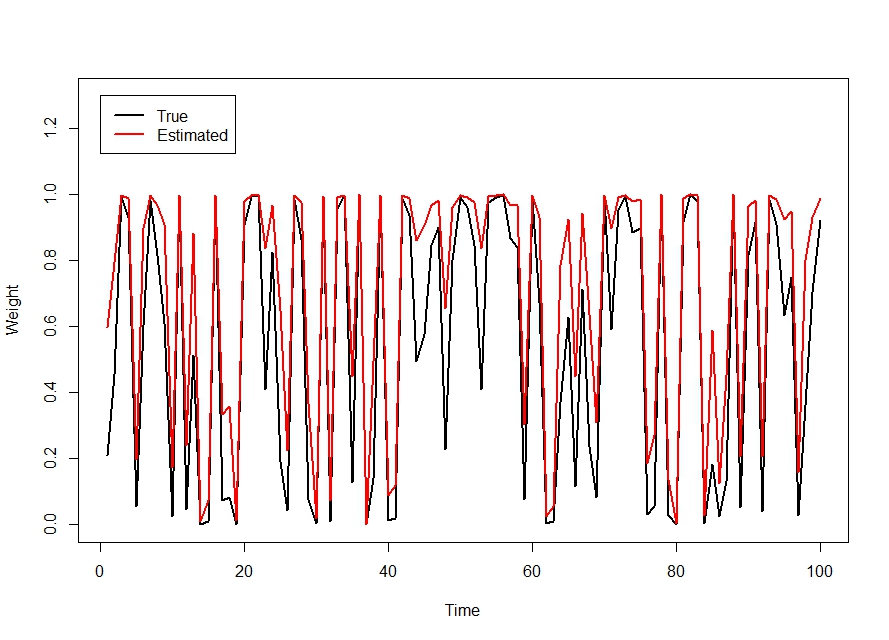}		
	\end{center}
	\caption{Weights estimation/prediction results when temporal correlation exists. Left plot: estimated weights under training data; right plot: predicted weights under test data.} 
	\label{fig:model4_p3}
\end{figure}

In this simulation, we set the parameters to be $p=2, \mu = \theta = 0, \sigma = \delta = 1, \sigma_a = 0.1, \phi = 0.5$ and $\pmb\beta = (1, 3, -2)$. Although the temporal correlation exists, when we apply our model 3 given in Eq. (\ref{eq:gcmixture3}), we obtain estimated parameters to be $\hat\mu = -0.013, \hat\sigma = 1.02, \hat\theta = -0.0019, \delta = 1.52$ and $\pmb{\hat\beta} = (2.52, 2.79, -1.72)$ which are reasonably close to the true values. In addition, the estimated/predicted $\hat\alpha_i$ can also capture the true dynamic pattern of the weights (see figure \ref{fig:model4_p3} for details). Therefore, our model 3 is robust even when temporal correlation exists, and thus we will only apply model 3 in the later real data analysis.

\section{Applications in Financial Data}

\subsection{Models without Exogenous Variables}
In this first real data application example, the data we use is the weekly stock prices of Citibank, Bank of America, JPMorgan Chase, and Dow Jones indexes from 1990 to 2014 (the data is downloaded from Yahoo finance). The corresponding asset log returns are computed for the later analysis.

\begin{figure}[H]
\begin{center}
\includegraphics[scale = 0.5]{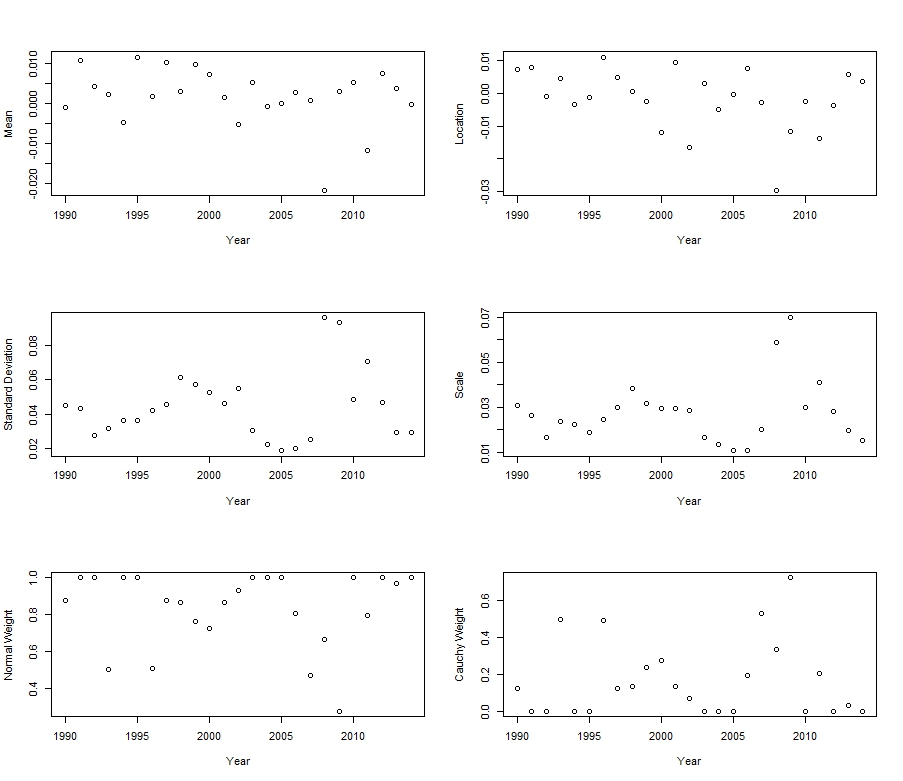}
\end{center}
\caption{Estimated parameters in Gaussian-Cauchy mixture model using asset return data of Citibank stock prices.}
\label{fig:gc_citi}
\end{figure}

In the first model, we assume that the returns in each year are independent with each other so that we can estimate all the parameters for each year individually. All of the four assets have very similar estimation results. Here we only use the results for Citibank as an illustration (see details in figure \ref{fig:gc_citi}). It is obvious that the estimated location and scale parameters of the two distributions are different through the year. The location parameters of both distributions for the year 2008 are negative and the corresponding scale parameters are huge, which matches with the terrible economic condition in the market that year. In addition, during this economically stressed period, say 2008--2011, the weights for Cauchy distribution become larger, with the largest value slightly less than 0.8.

We also apply our model 2 to the log returns of these four assets, that is, to treat the Gaussian and Cauchy distributions having constant parameters through the years, and only estimate the weights as time-varying variable. The estimation results of the distribution parameters are given in table \ref{table:model2}, from which we can observe that the location parameters of Gaussian distribution are larger than those of Cauchy distribution for all four assets, and the scale parameters of Gaussian distributions are smaller. This finding matches with the design of these two distributions, that is, Gaussian distribution represents the good economic condition, while Cauchy distribution stands for the bad economic condition.

\begin{table}[H]
\caption{Estimated parameters for Gaussian and Cauchy distributions from 1990 to 2014.}
\label{table:model2}
\vspace{8pt}
\centering
\begin{tabular}{cc|cccc}
	&		&	BOA	&	Citi Bank	&	JP Morgan	&	Dow Jones	\\\hline
\raisebox{-8pt}{Gaussian}	&	$\mu$	&	0.00319	&	0.00330	&	0.00287	&	0.00278	\\
	&	$\sigma$	&	0.02716	&	0.03417	&	0.03226	&	0.01791	\\\hline
\raisebox{-8pt}{Cauchy}	&	$\theta$	&	0.00154	&	-0.00372	&	0.00207	&	0.00273	\\
	&	$\delta$	&	0.03191	&	0.04555	&	0.04087	&	0.01987	\\
\end{tabular}
\end{table}

\begin{figure}
	\begin{center}
		\includegraphics[scale = 0.38]{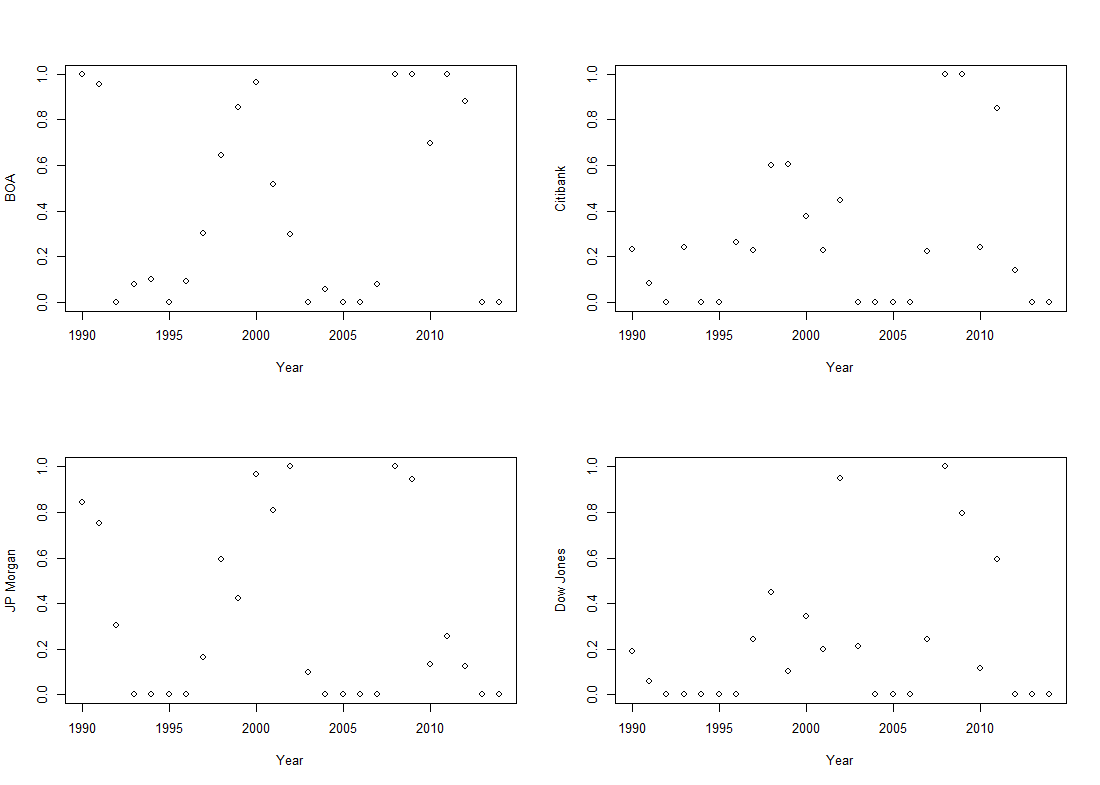}
	\end{center}
	\caption{Estimated time-varying weights for Cauchy distribution from 1990 to 2014.}
	\label{fig:model2_weights}
\end{figure}

The weight estimation results of model 2 are plotted in figure \ref{fig:model2_weights}, from which we can have a more clear idea about the market economic situation. For example, we can observe larger weights for Cauchy distribution for the ``dot-com'' bubble around 2000, and the 2008 financial crisis. For the year 2008, the weight for Cauchy distribution reaches to almost 1. Comparing to the results in model 1, we can find that the weight $\alpha_i$ in model 2 is a better indicator to the market volatility, because part of the volatility in model 1 is explained by the time-varying scale parameters of the two distributions. 

\subsection{Model with Exogenous Variables}
In the second real data example, we consider to utilize six exogenous variables as predictors to model the weight $\alpha_i$, which include consumer price index (CPI), producer price index (PPI), M1, unemployment rate, balance of trade (BOT) and housing starts. They are considered as significant factors that may affect the stock returns \citep{eckbo2000seasoned, flannery2002macroeconomic}. 

\begin{figure}[H]
	\begin{center}
		\includegraphics[scale = 0.4]{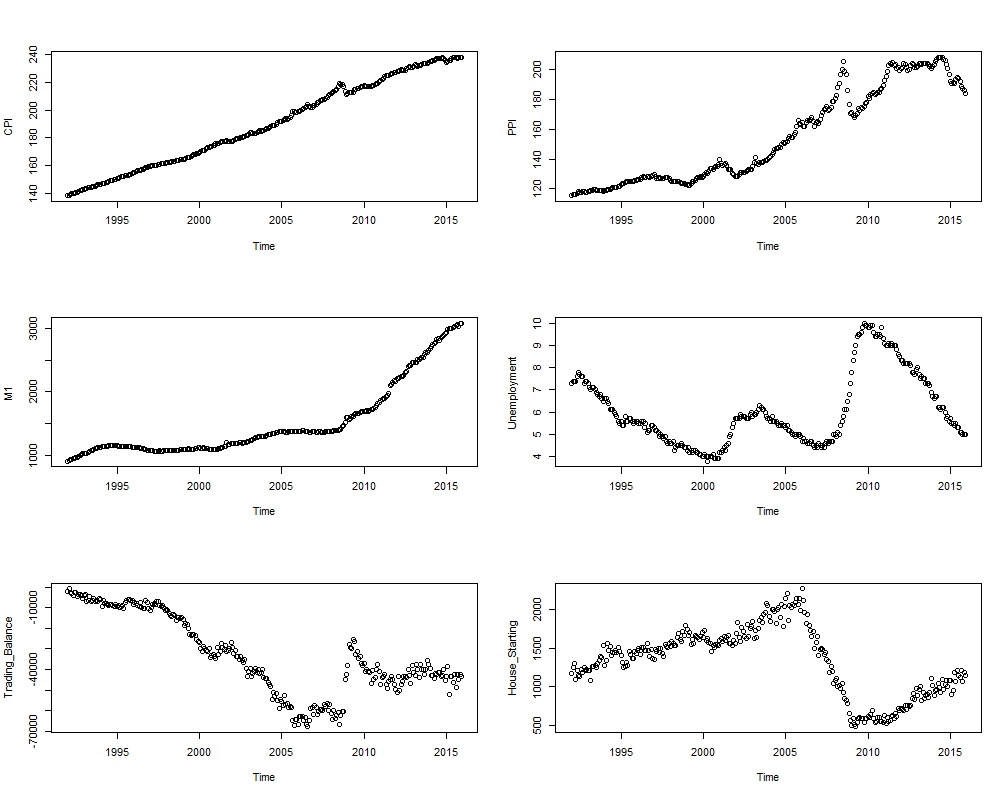}
	\end{center}
	\caption{Trend of six monthly macroeconomic indexes from 1992 to 2015.}
	\label{fig:eco_ind}
\end{figure}

The data we use are the monthly data from 1992 to 2015 with their time series plots given in figure \ref{fig:eco_ind}, where CPI, PPI, unemployment rate are downloaded from US. Bureau of Labor Statistics, M1 is obtained from Board of Governors of the Federal Reserve System, and BOT and housing starts are from US Census Bureau. Since there exists strong correlation (0.97) between CPI and PPI, in our analysis we only use CPI as one of the predictors. In addition, we use the first-order difference of CPI to stabilize its mean. The response variable used in this analysis is the daily compound return of $S\&P500$ index downloaded from Yahoo finance. We apply our model 3 to the data and the estimated parameters are shown in table \ref{table:model3_sp}.

\begin{figure}
	\begin{center}
		\includegraphics[scale = 0.5]{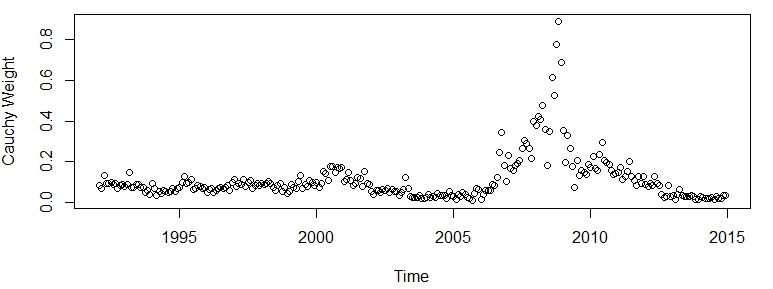}
	\end{center}
	\caption{Estimated weights of Cauchy distribution from 1992 to 2015.}
	\label{fig:m3_wt}
\end{figure}

In table \ref{table:model3_sp}, we can find that all predictors are significant. In addition, the location parameter for the Gaussian distribution is positive, 0.0012, which represents the good economic condition, while the one for the Cauchy distribution is negative, -0.014, representing for the loss in the bad economic condition. The behavior of weight for Cauchy distribution, $1-\alpha_i$, is given in figure \ref{fig:m3_wt}, from which we can find that (1) the weights within each year are actually not constant, especially during the financial crisis periods; and (2) the weight for Cauchy distribution becomes larger during the financial crisis, especially between 2008 and 2011. 

\begin{table}
	\caption{Parameter estimation results for the data with exogenous variables. The standard errors are computed by using Fisher information matrix.}
	\label{table:model3_sp}
	\vspace{5pt}
	\centering
	\begin{threeparttable}
		\begin{tabular}{ r|  c c  c  c  c c c}
			\hline
			Parameter &&& $\mu$ & $\sigma$ & $\theta$ &  $\delta$  \\[2pt]
			\hline
			Estimates &&& 0.0012 	& 0.0078  	& -0.014 	&  0.011 \\ [2pt]
			Std. Err.  &&& 0.00015	& 0.00014   & 0.0016	& 0.0010 \\[2pt]
			\hline 
			\multicolumn{8}{c}{} \\
			\hline
			Parameter & & Intercept & CPI & M1 &  Unemploy. & BOT &  Housing\\[2pt]
			\hline
			Estimates & & 2.37 	& 0.31  	& 0.99 	&  0.50	&  0.57	& 1.34\\ [2pt]
			Std. Err.  & & 0.13	& 0.059   & 0.14	& 0.12  & 0.080 & 0.15 \\[2pt]
			\hline
		\end{tabular}
	\end{threeparttable}
\end{table}

\section{Conclusion}
Financial risk management is a critical part of regular financial market operation. Although financial market data usually display leptokurtosis and skewness, Gaussian distribution is commonly used due to its easiness to implement. Since Gaussian distribution has a thin tail, it may fail to predict large losses. Hence in this paper, we propose a two-component mixture model which is a combination of Gaussian distribution and a fat-tail Cauchy distribution. In 
order to make our model to have the flexibility to adapt to different economic conditions, the weight of the Gaussian component can be changed over time.

We developed a series of four models---from simple to complex. In the first model, we assumed identical distribution within each time periods and independence between different time periods. With this setup, we can estimate parameters for each time period independently. However, it is more natural to assume  the location and scale parameters of two components are fixed over time while only let the weights change. This is how we did in the second model. In the third model, we incorporated a logistic model to predict the weights. Predictors to be used are macroeconomic indexes which could indicate the economic situations. In our final model, we considered the temporal correlation and added an AR error term in the logistic model. The estimation algorithms for all four models were established.   

We conducted simulation studies to check the performance of these models. In addition, we applied our models to the real data, where we can observe that the time-varying weight can indeed effectively indicate the economic situation. Around years 2000 and 2008, we can always observe spikes of weights for Cauchy distribution. This time-varying mixture model can also be applied to other financial market data and in other fields as long as the assumption holds that data are from two populations and the proportion changes over time.

\bibliographystyle{ims}
\bibliography{mix}
\appendix
\section{Appendix: Supplementary Algorithms and Tables}

\begin{algorithm}
\caption{Parameter Estimation for Gaussian-Cauchy Mixture Model}
\label{alg:gcmixture}
\begin{algorithmic}[1]
\Require{data vector $\mathbf{y}=(y_1, y_2, \dots, y_n)^T$}. 
\State Set the initial estimates
	\State $\mu_0 \gets median(\mathbf{y})$
	\State $\sigma_0 \gets IQR(\mathbf{y})$
	\State $\theta_0 \gets median(\mathbf{y})$
	\State $\delta_0 \gets IQR(\mathbf{y})$
	\State $\alpha_0 \gets 0.5$
\While {not converge}
	\State  $p_i \gets \frac{\alpha_0 f_g(y_i, \mu_0, \sigma_0)}{\alpha_0 f_g(y_i, \mu_0, \sigma_0)+(1-\alpha_0) f_c(y_i, \theta_0, \delta_0)}$ for $i = 1, 2, \dots, n$
	\State 	$\alpha_{new} \gets \frac{\sum\limits_{i=1}^{n}p_i}{n}$,\\
	\State	$\mu_{new} \gets \frac{\sum\limits_{i=1}^{n}p_i y_i}{\sum\limits_{i=1}^{n}p_i} $,\\
	\State	$\sigma_{new} \gets \sqrt{\frac{\sum\limits_{i=1}^{n}(y_i - \mu_{new})^2 p_i}{\sum\limits_{i=1}^{n}p_i}}$
	\State	$(\theta_{new}, \delta_{new}) \gets \argmax\limits_{\theta, \delta} \sum\limits_{i=1}^{n}\Big(\big(-\log\delta-\log(1+(\frac{y_i-\theta}{\delta})^2)\big)(1-p_i)\Big)$ \label{step:theta_delta}
	\State $\alpha_0 \gets \alpha_{new}, \mu_0 \gets \mu_{new}, \sigma_0 \gets \sigma_{new}, \theta_0 \gets \theta_{new}, \delta_0 \gets \delta_{new}$
\EndWhile \\
\Return par = $(\mu_0, \sigma_0, \theta_0, \delta_0, \alpha_0)$, 
\end{algorithmic}
\end{algorithm}


\begin{algorithm}
\caption{Parameter Estimation for Gaussian-Cauchy Mixture Model with Constant Location and Scale Parameters}
\label{alg:gcmixture2}
\begin{algorithmic}[1]
\Require{data vector $\mathbf{y}=(y_{11}, \dots, y_{1n_1}, y_{21}, \dots, y_{2n_2}, y_{k1},\dots, y_{kn_k})^T$}.
\State apply algorithm \ref{alg:gcmixture} to $\mathbf{y}$ to get estimated parameters $\mu_0, \sigma_0, \theta_0, \delta_0, \alpha_0$ as initial values
	\State $\alpha_{10} = \alpha_{20} = \dots = \alpha_{k0} \gets \alpha_0$
\While {not converge}
	\State  $p_{ij} \gets \frac{\alpha_{i0} f_g(y_{ij}, \mu_0, \sigma_0)}{\alpha_{i0} f_g(y_{ij}, \mu_0, \sigma_0)+(1-\alpha_{i0}) f_c(y_{ij}, \theta_0, \delta_0)}$ for $i = 1, 2, \dots, k$ and $j = 1, 2, \dots, n_i$
	\State 	$\alpha_{i, new} \gets \frac{\sum\limits_{j=1}^{n_i}p_{ij}}{n_i}$ for $i = 1, 2, \dots, k$\\
	\State	$\mu_{new} \gets \frac{\sum\limits_{i=1}^{k}\sum\limits_{j=1}^{n_i}p_{ij} y_{ij}}{\sum\limits_{i=1}^{k}\sum\limits_{j=1}^{n_i}p_{ij}} $,\\
	\State	$\sigma_{new} \gets \sqrt{\frac{\sum\limits_{i=1}^{k}\sum\limits_{j=1}^{n_i}(y_{ij} - \mu_{new})^2 p_{ij}}{\sum\limits_{i=1}^{k}\sum\limits_{j=1}^{n_i}p_{ij}}}$
	\State	$(\theta_{new}, \delta_{new}) \gets \argmax\limits_{\theta, \delta} \sum\limits_{i=1}^{k}\sum\limits_{j=1}^{n_i}\Big(\big(-\log\delta-\log(1+(\frac{y_{ij}-\theta}{\delta})^2)\big)(1-p_{ij})\Big)$ 
	\State $\alpha_{i0} \gets \alpha_{i, new}$ for $i = 1, 2, \dots, k$
	\State $\mu_0 \gets \mu_{new}, \sigma_0 \gets \sigma_{new}, \theta_0 \gets \theta_{new}, \delta_0 \gets \delta_{new}$
\EndWhile \\
\Return par = $(\mu_0, \sigma_0, \theta_0, \delta_0, \alpha_{10}, \alpha_{20}, \dots, \alpha_{k0})$.
\end{algorithmic}
\end{algorithm}

\begin{table}
	\caption{The detailed $\alpha$ estimation results for Gaussian-Cauchy mixture model defined in Eq. (\ref{eq:gcmixture2}).}
	\label{sim:m2gcmixture_extra}
	\vspace{5pt}
	\centering
	\begin{threeparttable}
		\begin{tabular}{ c | c  c  c  c }
			Parameter & True Value & Estimate Average & Estimate SE & MSE \\ [2pt]
			\hline 
			$\alpha_1$ & 0.7 	& 0.67 		& 0.14 	& 0.021 \\ [2pt]
			$\alpha_2$ & 0.75 	& 0.76 		& 0.13 	& 0.017 \\ [2pt]
			$\alpha_3$ & 0.8 	& 0.80 		& 0.14 	& 0.020 \\ [2pt]
			$\alpha_4$ & 0.85 	& 0.89 		& 0.09 	& 0.010 \\ [2pt]	 		
			$\alpha_5$ & 0.9 	& 0.91 		& 0.10 	& 0.010 \\ [2pt]
			$\alpha_6$ & 0.1 	& 0.20 		& 0.17 	& 0.039 \\ [2pt]			
			$\alpha_7$ & 0.2 	& 0.27 		& 0.19 	& 0.041 \\ [2pt]
			$\alpha_8$ & 0.6 	& 0.64 		& 0.16 	& 0.027 \\ [2pt]			
			$\alpha_9$ & 0.75 	& 0.78 		& 0.11 	& 0.013 \\ [2pt]
			$\alpha_{10}$ & 0.9 	& 0.90 		& 0.10 	& 0.010 \\ [2pt]			
			\hline
		\end{tabular}
	\end{threeparttable}
\end{table}
\end{document}